\documentclass{emulateapj}

\def\ha{H$\alpha$}
\def\hb{H$\beta$}

\def\oii{[OII]$\lambda$3727}

\def\oiiib{[OIII]$\lambda$5007}
\def\oiii{[OIII]$\lambda$4958,5007}

\newcommand {\mean} [1] {\langle#1\rangle}

\shorttitle{Both `negative' and `positive' feedback in an obscured high-z Quasar}
\shortauthors{Cresci et al.}

\begin{document}


\title{Blowin' in the wind: both `negative' and `positive' feedback in an obscured high-z Quasar}\altaffiltext{$\star$}{SINFONI data were obtained from observations made with the ESO Telescopes at the Paranal Observatories (ESO programs 092.A-0144 and 383.A-0573).}


\author{
G. Cresci\altaffilmark{1},\email{gcresci@arcetri.astro.it} V. Mainieri\altaffilmark{2}, M. Brusa\altaffilmark{3,4,5}, A. Marconi\altaffilmark{6}, M. Perna\altaffilmark{3}, F. Mannucci\altaffilmark{1}, E. Piconcelli\altaffilmark{7}, R. Maiolino\altaffilmark{8,9}, C. Feruglio\altaffilmark{7,10}, F. Fiore\altaffilmark{7}, A. Bongiorno\altaffilmark{7}, G. Lanzuisi\altaffilmark{3,5}, A Merloni\altaffilmark{4}, M.Schramm\altaffilmark{11}, J.~D. Silverman\altaffilmark{11}, F. Civano\altaffilmark{12}
}

\altaffiltext{1}{INAF - Osservatorio Astrofisco di Arcetri, largo E. Fermi 5, 50127, Firenze, Italy}
\altaffiltext{2}{European Southern Observatory, Karl-Schwarzschild-str. 2,  85748 Garching bei M\"unchen, Germany}
\altaffiltext{3}{Dipartimento di Fisica e Astronomia, Universit\`a di Bologna, viale Berti Pichat 6/2, 40127 Bologna, Italy}
\altaffiltext{4}{Max Planck Institut f\"ur Extraterrestrische Physik, Giessenbachstrasse 1, 85748 Garching bei M\"unchen, Germany}
\altaffiltext{5}{INAF - Osservatorio Astronomico di Bologna, via Ranzani 1, 40127 Bologna, Italy}
\altaffiltext{6}{Universit\`a degli Studi di Firenze, Dipartimento di Fisica e Astronomia, via G. Sansone 1, 50019 Sesto F.no, Firenze, Italy}
\altaffiltext{7}{INAF - Osservatorio Astronomico di Roma, via Frascati 33, 00040 Monteporzio Catone, Italy}
\altaffiltext{8}{Cavendish Laboratory, University of Cambridge, 19 J.J. Thomson Ave., Cambridge, UK}
\altaffiltext{9}{Kavli Institute for Cosmology, University of Cambridge, Madingley Road, Cambridge, UK}
\altaffiltext{10}{IRAM - Institut de RadioAstronomie Millim\'etrique, 300 rue de la Piscine, 38406 Saint Martin d’H\`eres, France}
\altaffiltext{11}{Kavli Institute for the Physics and Mathematics of the Universe (WPI), Todai Institutes for Advanced Study, the University of Tokyo, Kashiwanoha 5-1-5, Kashiwa-shi, Chiba 277-8568, Japan}
\altaffiltext{12}{Department of Physics and Yale Center for Astronomy and Astrophysics, Yale University, P.O. Box 208121, New Haven, CT 06520-8121, USA}


\begin{abstract}
Quasar feedback in the form of powerful outflows is invoked as a key mechanism to quench star formation in galaxies, preventing massive galaxies to over-grow and producing the red colors of ellipticals. On the other hand, some models are also requiring `positive' AGN feedback, inducing star formation in the host galaxy through enhanced gas pressure in the interstellar medium. However, finding observational evidence of the effects of both types of feedback is still one of the main challenges of extragalactic astronomy, as few observations of energetic and extended radiatively-driven winds are available.
Here we present SINFONI near infrared integral field spectroscopy of XID2028, an obscured, radio-quiet $z=1.59$ QSO detected in the XMM-COSMOS survey, in which we clearly resolve a fast (1500 km/s) and extended (up to 13 kpc from the black hole) outflow in the [OIII] lines emitting gas, whose large velocity and outflow rate are not sustainable by star formation only. 
The narrow component of H$\alpha$ emission and the rest frame U band flux from HST-ACS imaging enable to map the current star formation in the host galaxy: both tracers independently show that the outflow position lies in the center of an empty cavity surrounded by star forming regions on its edge. The outflow is therefore removing the gas from the host galaxy (`negative feedback'), but also triggering star formation by outflow induced pressure at the edges (`positive feedback'). XID2028 represents the first example of a host galaxy showing both types of feedback simultaneously at work.
\end{abstract}


\section{Introduction}

Outflows driven by Active Galactic Nuclei (AGN) are expected to sweep away most of the gas in their host galaxy, hence quenching both star formation and further black hole accretion, yielding to the black hole-galaxy mass relation observed locally (e.g., Granato et al. \citealp{granato04}, Di Matteo et al. \citealp{dimatteo05}, Menci et al. \citealp{menci06}). Quasar feedback is also thought to be the main mechanism preventing massive galaxies to over-grow (explaining the dearth of very massive galaxies) and responsible for the red colors of ellipticals. On the other hand, some models are also requiring `positive' AGN feedback, inducing star formation in the host galaxy through enhancing the gas turbulence in the interstellar medium, to reproduce the observed correlation between nuclear star forming activity and AGN luminosity, the $M_{BH}-\sigma$ scaling relation and the enhanced specific star formation rate ($sSFR=SFR/M_*$) at high redshift (e.g. Silk \citealp{silk13}, King \citealp{king05}, Ishibashi \& Fabian \citealp{ishibashi12}). However, finding observational evidence of the effects of such quasar `negative' and `positive' feedback is still one of the main challenges of extragalactic astronomy.

According to AGN-galaxy co-evolutionary models, the most luminous sources  (L$_{\rm bol}>10^{46}$ erg s$^{-1}$) experience the so-called ‘feedback’ or ‘blow-out’ phase after an early dust enshrouded phase associated with rapid SMBH growth and violent star formation episodes (e.g. Menci et al. \citealp{menci08}; Hopkins et al. \citealp{hopkins08}, Lapi et al. \citealp{lapi14}). Shortly later, the accreting BH releases the energy necessary to heat or expel the gas, in the form of outflowing winds, manifesting itself as an X-ray and visible quasar.

While powerful outflows sustained by kinetic energy ejected in the hosts of luminous radio-galaxies by relativistic and collimated jets have been commonly detected out to z$\sim4$ (see e.g. Nesvadba et al. \citealp{nesvadba08}, \citealp{nesvadba10}),  radiatively-driven winds are less commonly observed, and only very recently spatially resolved optical, IR and mm spectroscopic studies convincingly showed the first evidences for the existence of such processes  at low-z  (e.g. Feruglio et al. \citealp{feruglio10}, Rupke \& Veilleux \citealp{rupke11}, Sturm et al. \citealp{sturm11}, Arav et al. \citealp{arav13}, Feruglio et al. \citealp{feruglio13}, Cicone et al. \citealp{cicone14}) and at high-z, both in luminous QSOs (Maiolino et al. \citealp{maiolino12}, Harrison et al. \citealp{harrison12}, Cano-D\'iaz et al. \citealp{cano12}) and massive star forming galaxies (F\"orster-Schreiber et al. \citealp{natascha14}).
However at high-z, where the effects of feedback are expected to be more effective, most of the results are still based on the observations of very luminous, unobscured QSOs, in which the blow-out phase is expected to be near its end. Instead, if large scale, energetic radiatively driven winds are rare because they arise during a short-lived stage ($<100$ Myr), the crucial point is to select objects at the maximum of the `feedback' phase, when the obscuration is still significant. 

We have recently completed two follow-up programs on obscured quasars pre-selected for being in a significant outflowing phase. The first consists in X-Shooter observations of a sample of X-ray selected obscured QSOs at z$\sim1.5$, selected from the XMM-COSMOS survey (Hasinger et al. \citealp{hasinger07}) on the basis of their observed red colors (R-K$>4.5$) and high MIR and X-ray to optical flux ratios (X/O$>10$, see Brusa et al. \citealp{brusa10}). The presence of outflowing material identified by \oiii\ kinematics was indeed detected in 6 out of 8 sources, confirming the efficiency of a selection based on X-ray to optical to NIR colors in isolating such objects undergoing a `blow-out' phase (Brusa et al. \citealp{brusa14}). Moreover, in the two brightest targets (XID2028 and XID5321) we were able to perform slit-resolved spectroscopy and detect an extended wing of \oiiib\ on scales of $\sim$6-10 kpc from the central AGN, suggesting the presence of a massive, galactic-scale outflow (see Perna et al. \citealp{perna14}).

We also recently completed a SINFONI program on a sample of AGN pre-selected to have high accretion rates, where radiatively-driven winds are supposed to be more effective since they likely originate from the acceleration of disk outflows by the AGN radiation field, and a combination of moderate obscuration observed in the X-rays and high Eddington ratio where outflows or transient absorption are expected to happen (Fabian et al. \citealp{fabian08}).

Interestingly, XID2028 at $z=1.59$ was selected as an obscured, outflowing QSOs using both criteria described above. Moreover, it presents additional, indirect evidences of being an object caught in the feedback phase. It shows a point like nucleus embedded in an extended galaxy with asymmetric/disturbed morphology as witnessed by rest-frame U-band HST/ACS (Advanced Camera for Survey) data, and as expected in mergers models. The UV to FIR Spectral Energy Distribution (SED) is best fitted by the Mrk231 template, a local AGN where a powerful AGN driven outflow have been detected both in the mm (Feruglio et al. \citealp{feruglio10}) and in the FIR (Fischer et al. \citealp{fischer10}) wavelengths. The total bolometric AGN luminosity of XID2028, as determined from the SED decomposition, is L$_{\rm AGN}\sim2\cdot10^{46}\ erg\ s^{-1}$ (Lusso et al. \citealp{lusso12}). Despite the lack of any significant broad feature in the UV regime, it shows a very significant broad component in the H$\alpha$ line (FWHM$\sim5400$ km s$^{-1}$, from a detailed multi-component fit of the X-Shooter spectrum in Bongiorno et al. \citealp{bongiorno14}). Assuming virial arguments, this translates into a black hole mass $M_{\rm BH}\sim 2.7\cdot10^{9}\ M_{\odot}$. Such massive BH is hosted in a massive galaxy ($M_*\sim10^{12}\ M_{\odot}$ from SED decomposition fitting) ongoing substantial star formation ($SFR\sim275\ M_{\odot}/yr$ from PACS/Herschel data, see Brusa et al. \citealp{brusa14}), consistent with the level observed in main sequence galaxies of comparable mass at z$\sim1.5$ (Whitaker et al.\citealp{whitaker12}).

The target is detected in the radio band at $1.4\ GHz$ in the Very Large Array (VLA) observations of the COSMOS field (Schinnerer et al. \citealp{schinnerer07}). The radio flux is $102\pm20\ \mu Jy$ and the radio power implied by the faint detection ($L_{1.4} \sim 10^{24}\ W Hz^{-1}$) places this source in the radio quiet class. In fact, we derive a value for q$_{24}=1.36$, computed between the observed 24 micron and 1.4 GHz flux densities. This value places the source well outside the radio loud class (Bonzini et al. \citealp{bonzini13}), differently from most previous studies with IFU follow-up of bright quasars (e.g. Nesvadba et al. \citealp{nesvadba08},\citealp{nesvadba10}; Harrison et al. \citealp{harrison12}). 
Although we cannot completely exclude the contribution of a faint radio jet in this system, as observed in Mk231 by Rupke et al. \citep{rupke11}, the observed low level of radio emission usually implies negligible or very marginal contribution. In fact, even if radio jets may be important for shaping the properties of the extended narrow line region even in radio-quiet QSOs, they are usually not kinematically relevant at low radio luminosities (e.g. Husemann et al. \citealp{husemann13}).

In this paper we present SINFONI integral field spectroscopic observations of XID2028, needed to fully map the 2D outflow distribution using the [OIII]$\lambda$5007 blue-shifted wing, and compare the results with archive SINFONI narrow H$\alpha$ observations tracing the star formation in the host, and high spatial resolution images from HST. 
The paper is organized as follows: in Sect.~\ref{observations} we present the observations and data reduction, the evidences for a massive outflow from the J band datacube are presented in Sect.~\ref{Jresults}. The outflow energetics is discussed in Sect.~\ref{energetic}, while the outflow effects on the host galaxy derived from the H+K datacube as well as from HST-ACS imaging are discussed in Sect.~\ref{Hresults}. We present our conclusions in Sect.~\ref{conclusions}.






\section{Observations, Data Reduction and Analysis} \label{observations}

The galaxy XID2028 was observed with the near-IR Integral Field spectrograph SINFONI (Eisenhauer et al. \citealp{frank03}) in J band at the VLT UT4 telescope in the framework of program 092.A-0144 (PI Mainieri). The observations were executed in four different nights during February and March 2014. The J band grating was used to sample the redshifted [OIII]$\lambda$$\lambda$5007,4959 and H$\beta$ emissions, providing a spectral resolution of $R=2000$. The observations were carried out in seeing limited mode, using the $0.250" \times 0.125"$ pixel scale, which provides a total field of view of $8" \times 8"$. The average spatial resolution obtained, as sampled by combined Point Spread Function (PSF) reference star observations before and after every hour of target integration, is $\sim 0.7"$. During each Observing Block (OB), an ABBA nodding was used, putting the target in two positions of the field of view about $3.5"$ apart, with an integration time of 600 sec in each position. The total integration time was 6 hours on target.

SINFONI H+K observations of the source taken in May 2009 were available in the ESO archive (program 383.A-0573, PI McMahon). The H+K grating samples the H$\alpha$ and [NII]$\lambda$$\lambda$6548,6584 with $R=1500$. The data were obtained in seeing limited mode with the $0.250" \times 0.125"$ pixel scale, in four integrations of $300\ sec$ each, following an ABBA pattern, for a total of 20' exposure. A dedicated PSF star observation is not available, but the resolution obtained on the telluric star observed after the science exposure for flux calibration suggests a spatial resolution of $\sim0.9"$.

Both the J band and H+K data were reduced using the ESO pipeline (Modigliani et al. \citealp{modigliani07}), with the improved sky subtraction proposed by Davies et al. \citep{davies07}. After flat-fielding, sky subtraction, correction for distortions, cosmic rays removal and wavelength calibration, the two-dimensional (2D) data of each OB are mapped into a 3D data cube with dimensions $32 \times 64 \times 2048$ pixels. Telluric O/B type stars observations are used to flux calibrate each OB. The flux calibrated cubes of the different OBs are then combined together by measuring the relative offsets from the detected centroid of emission in a given spectral channel.  The final result of the data reduction procedure is a flux-calibrated datacube, containing the full image of the galaxy observed at each wavelength. 
We are thus able to extract from the cube both the one-dimensional spectrum at each pixel (or integrated over a larger region) and monochromatic images of the field of view at different wavelengths. 

Finally, XID2028 lies in the Cosmic Evolution Survey field (COSMOS\citep{scoville07}), and therefore benefits from deep HST/ACS coverage in the F814W filter  (1 orbit, corresponding to $\sim2000$s), achieving a limiting point-source 5$\sigma$ depth of AB=27.2 and a spatial scale of 0.05$"$/pixel (Koekemoer et al. \citealp{koekemoer07}). This filter samples the rest-frame U-band emission at z$\sim1.6$. 

In the following we will analyze the SINFONI J and H+K datacubes, fitting the line profiles on pixel to pixel basis for the brightest lines to measure and map the line emission distributions and the corresponding velocities. Given the highly asymmetric shape of the line profiles for \hb\ and \oiii, we have fitted separately the spectrum of each spatial spaxel in the field of view using a broken power law distribution convolved with a Gaussian: 
\begin{equation} \label{pareto}
	F_{\lambda}=\left\{\begin{array}{ll}
		F_0\times(\frac{\lambda}{\lambda_0})^{+\alpha} & \textrm{for $\lambda < \lambda_0$} \\
		F_0\times(\frac{\lambda}{\lambda_0})^{-\beta} & \textrm{for $\lambda > \lambda_0$}
\end{array} \right .
\end{equation}
The free parameters of the fit are, for each line, the central wavelength $\lambda_0$, the two power-law indices $\alpha$ and $\beta$, the Gaussian width $\sigma$ used for the convolution, and the normalization $F_0$. This function allows to reproduce asymmetric velocity profiles, and it is often used to fit QSO broad line regions (e.g. Nagao et al. \citealp{nagao06}). We also add a broad Gaussian component to reproduce the contribution of the BLR to the H$\beta$ line.
This choice reproduces well the complex line profile in our data, as shown in Fig.~\ref{Jband} for the $1"\times1"$ nuclear region, and has the advantage of having a single component for the [OIII] lines, with respect to multiple Gaussians, making it a more stable choice to fit the variation of the line profile across the field of view in the SINFONI datacube.

The \ha\ profile in the H+K data cube is instead dominated by the broad component due to the BLR, considerably extincted in the \hb\ region, and that is well reproduced and fitted by a single, symmetric Gaussian. Lower S/N residuals are present, and they are discussed in Sect.~\ref{Hresults}.


\section{Tracing the outflow with [OIII]$\lambda$5007} \label{Jresults}

\begin{figure*}
	\begin{center}
		\includegraphics[width=0.7\textwidth]{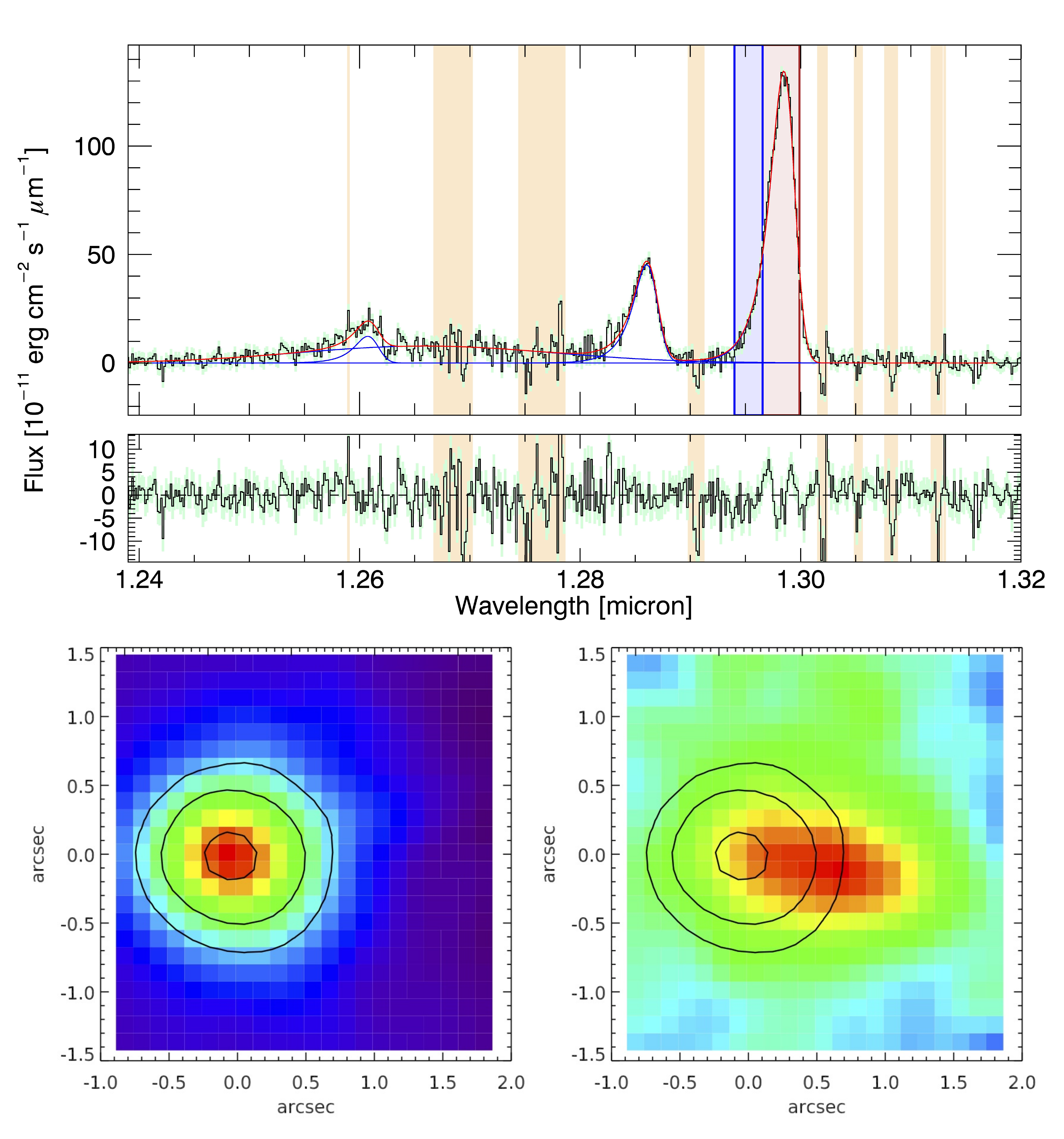}
	\end{center}
	\caption{[OIII] in XID2028. \textit{Upper panels:} the J band SINFONI spectrum of XID2028, integrated in a region of 8$\times$8 spaxels ($1"\times 1"$) around the QSO. The observed spectrum is shown in black, the different broken power-law components in the fit for each line (H$\beta$, large BLR H$\beta$, [OIII]$\lambda$$\lambda$4959,5007) are shown in blue, while their sum is shown in red. The shaded regions show the location of sky lines, that were excluded from the fit. The red and blue box show the intervals in wavelength in which the maps shown in the lower panels are integrated. The residuals of the fit, i.e. the difference between the observed and the model spectrum, are shown below. \textit{Lower Panels:} [OIII]$\lambda$5007 channel maps obtained integrating the continuum subtracted SINFONI datacube on the line core ($1.296<\lambda<1.300\ \mu m$, \textit{left}, see red box) and on the blue wing ($1.294<\lambda<1.296\ \mu m$, \textit{right}, see blue box). The contours on the line core (levels 0.3, 0.5, 0.9 relative to the peak), marking the position of the central QSO, are shown in black in both panels. The fully resolved, extended blue wing due to the outflow is extended up to $1.5"$, i.e. $13$ projected kpc from the QSO position. North is up and East is left, $1"$ corresponds to 8.5 kpc.} 
	\label{Jband}
\end{figure*}

The integrated  J band spectrum of XID2028  is shown in Fig.~\ref{Jband}: an asymmetric, prominent blue wing is clearly visible for all the emission lines, already suggesting the presence of outflowing material towards the observer. In fact, the forbidden [OIII] emission lines are an ideal tracer of extended outflowing ionized gas, as cannot be produced in the high-density, sub-parsec scales typical of AGN Broad-Line Regions (BLR). Therefore, an asymmetric, broad [OIII] profile has been interpreted as evidence of outflowing gas, both in AGN and in starburst dominated galaxies (see e.g. Harrison et al. \citealp{harrison12}, Mullaney et al. \citealp{mullaney13}). As often reported in such systems, a corresponding redshifted component is missing in XID2028, probably due to dust obscuration in the receding side of the outflow. 

To assess the spatial extent of the blueshifted wing, in Fig.~\ref{Jband} we also show the continuum subtracted SINFONI datacube collapsed on the spectral channels corresponding to the line core and to the line wing in the left and right panel respectively (lower panels). The prominence of an extended, fully resolved outflow is evident from the image of the blue wing in the west-east direction. The blue wing is originating on the QSO position, as marked from the line core black contours, and it is extending up to $1.5"$, i.e. 13 projected kpc, from the center.

\begin{figure*}
	\begin{center}
		\includegraphics[width=0.6\textwidth]{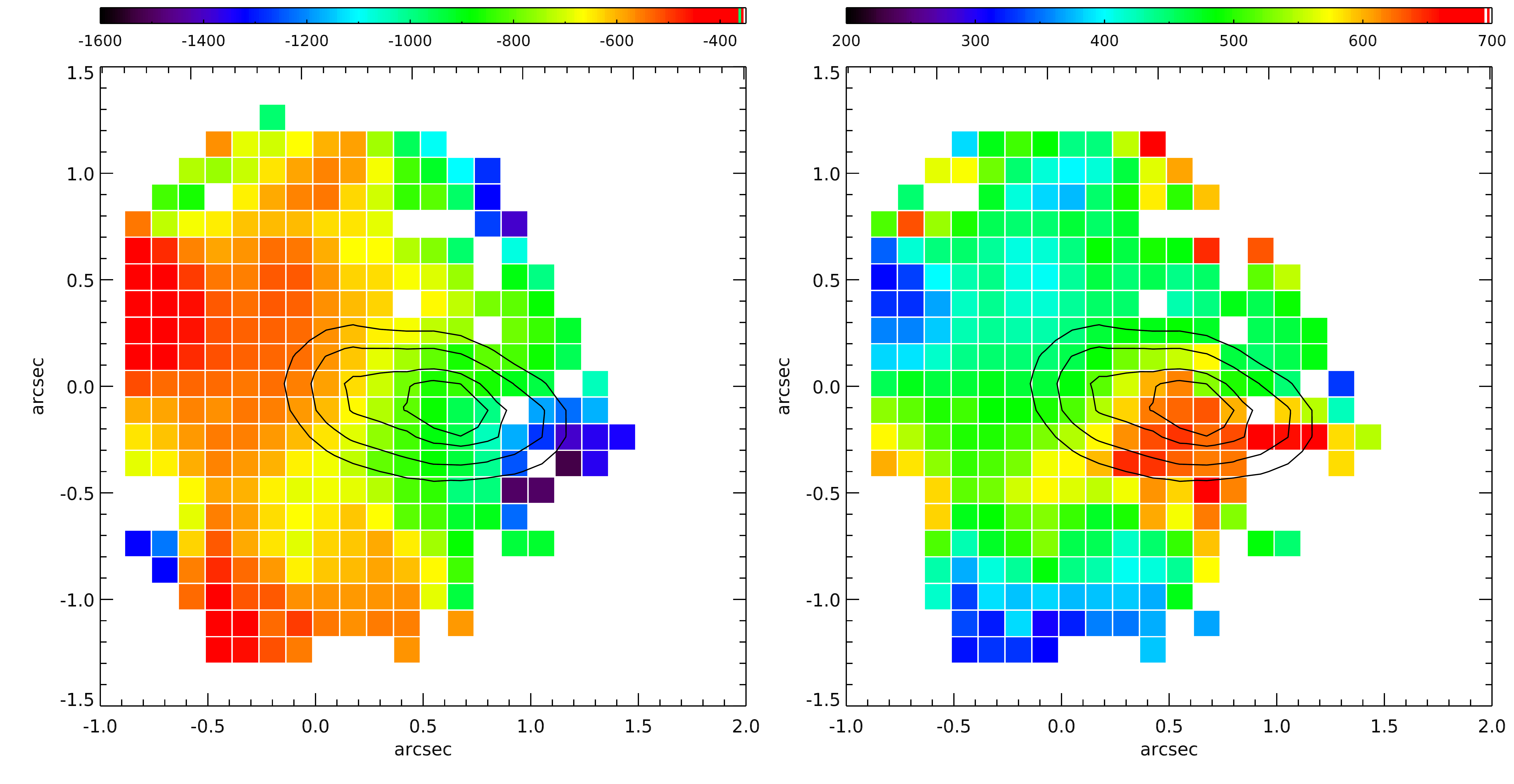}
	\end{center}
	\caption{Kinematics from the [OIII]$\lambda$5007 emission line profile. The \textit{left panel} shows the map of $v_{10}$, the velocity at the 10th percentile of the overall emission-line profile in each pixel. Velocities up to 1500 $km/s$ towards the line of sight are reached in the outflow region. The \textit{right panel} shows the width $W_{40}$, i.e. the velocity width of the line that contains 40\% of the emission line flux such that $W_{40}=v_{50}-v_{10}$, where $v_{50}$ and $v_{10}$ are the velocities at the 50th and 10th percentiles, respectively. 
The larger line widths ($W_{40}\sim600\ km/s$) are observed in the same regions where the strongly blueshifted gas is detected, as expected in the case of outflowing material. 
The blue wings contours from Fig.\ref{Jband} are superimposed in black (contour levels 0.7, 0.8, 0.9, 0.95 relative to the peak). North is up and East is left, $1"$ corresponds to 8.5 kpc.}
	\label{outvel}
\end{figure*}

We have 
fitted separately the spectrum of [OIII] in each spatial spaxel in the field of view using eq.~\ref{pareto}, in order to map the velocity of the line wing. Given the complexity of the line profile, we adopted a non-parametric definition to characterise the widths and velocities within each spaxel of the SINFONI data (see also Harrison et al. \citealp{harrison14}). In particular, a map across the field of view of $v_{10}$, the velocity at the 10th percentile of the overall emission-line profile in each pixel, is shown in the left panel of Fig.~\ref{outvel}, with the contours of the flux integrated over the blue wing (see Fig.~\ref{Jband}, lower right panel), and shows strongly blueshifted velocities in the outflow region, with velocities as high as $v_{10}=-1500\ km/s$ towards our line of sight.
The detected velocity are far too high to be due to rotational motions in the host galaxy. Moreover, in case of ordered rotation we expect the line width to be peaked at the center of the galaxy, where the velocity gradient is steeper and smeared out by the observational beam (e.g. Cresci et al. \citealp{cresci09}), while in this case the line width is larger in the blue wing region, supporting the interpretation that the high velocity is due to an outflow and not to kinematics in the host galaxy. 

We plot in the right panel of Fig.~\ref{outvel} the velocity width $W_{40}$, i.e. the width of the line that contains 40\% of the emission line flux such that $W_{40}=v_{50}-v_{10}$, where $v_{50}$ and $v_{10}$ are the velocities at the 50th and 10th percentiles, respectively. This is an esimate of the width of the blue wing of the line. The larger wing widths ($W_{40}\sim600\ km/s$) are observed in the same regions where the strongly blueshifted gas is detected, as expected in the case of outflowing material (M\"uller-Sanchez et al. \citealp{francisco11}). Similarly, small bulk velocity ($<200\ km/s$), velocity dispersions and distinct kinematic components have been observed in the few cases were galactic inflows have been detected, mostly in absorption (e.g. Bouch\'e et al. \citealp{bouche13}). Theoretical modelling also predicts that inflows should have small covering factor (Steidel et al. \citealp{steidel10}), explaining why direct observations of infalling gas have been so elusive and mostly indirect evidence of gas infall at high redshift are available (e.g. Cresci et al. \citealp{cresci10}). Finally, we rule out that the observed blueshifted velocities may be due to a merging galaxy, as the deep ACS image of the source do not show any counterpart at the blue wing peak (see Fig.~\ref{Hintegrated}), and this scenario would not explain the large line widths observed all over the blueshifted region.


\section{Energetic of the outflow and mass outflow rate} \label{energetic}

We derived the physical properties of the outflowing gas from the observed line emissions adopting a simple conical (or bi-conical) outflow distribution uniformly filled with outflowing clouds (see also Maiolino et al. \citealp{maiolino12}). 
We first estimate the outflowing ionized emitting gas from the \hb\ emission in the outflow region (Liu et al. \citealp{liu13}, Osterbrock \& Ferland \citealp{osterbrock06}), assuming $T_e=20000\ K$, as:
\begin{equation} 
	M_{ion}=2.82\cdot10^9\ M_{\odot}\ \left(\frac{L_{H_{\beta}}}{10^{43}\ erg\ s^{-1}}\right) \left(\frac{n_e}{100\ cm^{-3}}\right)^{-1}
\end{equation}
Assuming $n_e=100\ cm^{-3}$, and using the \hb\ luminosity $L_{H_{\beta}}=3\cdot10^{42}\ erg\ s^{-1}$ measured on the outflow region after removing the continuum and the broad component due to the BLR (see Fig.~\ref{Jband}), with no extinction correction, we get:
\begin{equation} \label{massout}
	M_{ion} \gtrsim 8.5\cdot10^8\ M_{\odot}
\end{equation}
We note that this is a lower limit given the unknown correction due to the dust extinction in the outflow, and that the uncertainty on the measure of $L_{H\beta}$ is much small compared to all our assumptions. The maximum outflow velocity inferred from our kinematic analysis is $v_{out}\sim1500\ km/s$. We assume in the following that this is representative of the average outflow velocity, i.e. the de-projected velocity of the outflow, and regard the lower velocities observed due to projection effects. This is again a lower limit if the (bi-)conical outflow does not intercept the line of sight. Given the radius derived for the outflow $R_{out}=13\ kpc$, the dynamical time is given by:
\begin{equation} \label{tdyn}
	t_d \approx R_{out}/v_{out} \sim 8.5\ Myr 
\end{equation}
which is consistent with similar outflows reported in the literature (Greene et al. \citealp{greene12}), as well as with the expected typical AGN lifetime (Martini et al. \citealp{martini01}).  

Given our hypothesis of a (bi-)conical outflow distribution out to a radius $R_{out}$ for the ionized wind, uniformly filled with outflowing clouds, the volume-averages density of the gas is:
\begin{equation}
	\mean{\rho_{out}}_V=\frac{M_{out}}{\Omega/3\cdot R^3_{out}}	
\end{equation}
where $\Omega$ is the solid angle subtended by the (bi-)conical outflow. The mass outflow rate is therefore given by:
\begin{equation}
	\dot M_{out} \approx \mean{\rho_{out}}_V \cdot \Omega R^2_{out} \cdot v_{out} = 3\ v_{out}\ \frac{M_{out}}{R_{out}}
\end{equation}
Note that the outflow rate is actually independent of both the opening angle $\Omega$ of the outflow and of the filling factor $f$ of the emitting clouds (under the assumption of clouds with the same density). 
Using the outflowing mass derived in eq.~\ref{massout} we obtain $\dot M_{out,ion}>300\ M_{\odot}/yr$. This represents an estimate for the ionized component only: if the QSO is also driving a neutral/molecular outflow, the total mass outflow rate is probably up to an order of magnitude larger (e.g. scaling by the same neutral-to-ionized fraction as in local QSOs, e.g. Rupke et al. \citealp{rupke13}).  In fact, evidences for the presence of a neutral outflow are derived from NaID absorption lines in the X-Shooter spectra (see Perna et al. \citealp{perna14}). Therefore derive that:
\begin{equation} \label{massrate}
	\dot M_{out}\gtrsim 1000\ M_{\odot}/yr
\end{equation}
although further observations at mm wavelengths sampling the molecular gas phase are required to assess the total mass and energy output, given the several assumptions and limits in this analysis. The corresponding kinetic power is given by:
\begin{equation} \label{kinpow}
	P_{kin,tot}\gtrsim \frac{1}{2}\ \dot M_{out}\ v_{out}^2 = 5.3\cdot 10^{44}\ erg\ s^{-1}
\end{equation}
The momentum flux (i.e. $\dot P=\dot M_{out} \times v_{out}$) inferred for this large-scale outflow is $9.5\cdot10^{36}\ dyne$, which corresponds to an outflow momentum rate relative to $L_{AGN}/c$ (i.e. the ``momentum boost'') of $\sim$ 10. This corresponds to what is expected in case of an energy-conserving outflows, where the momentum boost is linked to the velocity of the fast wind originated from the immediate vicinity of the AGN accretion disk that triggers the acceleration of  the galactic-scale, massive outflow (Faucher-Gigu{\`e}re et al. \citealp{fg12}). In fact, assuming a nuclear wind velocity of 0.1$c$ (as typically estimated for X-ray ultra-fast outflows by X-ray observations, e.g. King et al. \citealp{king10}, Tombesi et al. \citealp{tombesi10}), for a velocity of $1500\ km/s$ we would expect a momentum boost of $\sim$10. 

Such high velocities, mass outflow rate and kinetic power cannot be sustained by star formation only. In fact, the derived limit on the ratio between the mass outflow rate and the SFR for XID2028 is $\dot M_{out}/SFR > 3$, while it has been measured to be $\sim1$ in star forming driven outflow (e.g. Cicone et al. \citealp{cicone14}). Accordingly, the kinetic output expected from supernovae and stellar winds (Veilleux et al. \citealp{veilleux05}) for the star formation in XID2028, $P(SF)\sim7\cdot10^{41} \times SFR(M_{\odot}/yr)$, is at least 2.5 times lower than the measured kinetic power. Moreover, wind velocities above $1000\ km/s$ are extremely difficult to attain in starburst driven winds (see Thacker et al. \citealp{thacker06}, Lagos et al. \citealp{lagos13}). Therefore the central QSO seems to be the engine driving such a powerful outflow, as also suggested by the geometry of the outflowing gas, with the apparent starting point coincident with the AGN position in the nucleus of the galaxy (see Fig.~\ref{Jband}) .


\section{Effects of the outflow on the host galaxy} \label{Hresults}

\begin{figure}
	\begin{center}
		\includegraphics[width=0.5\textwidth]{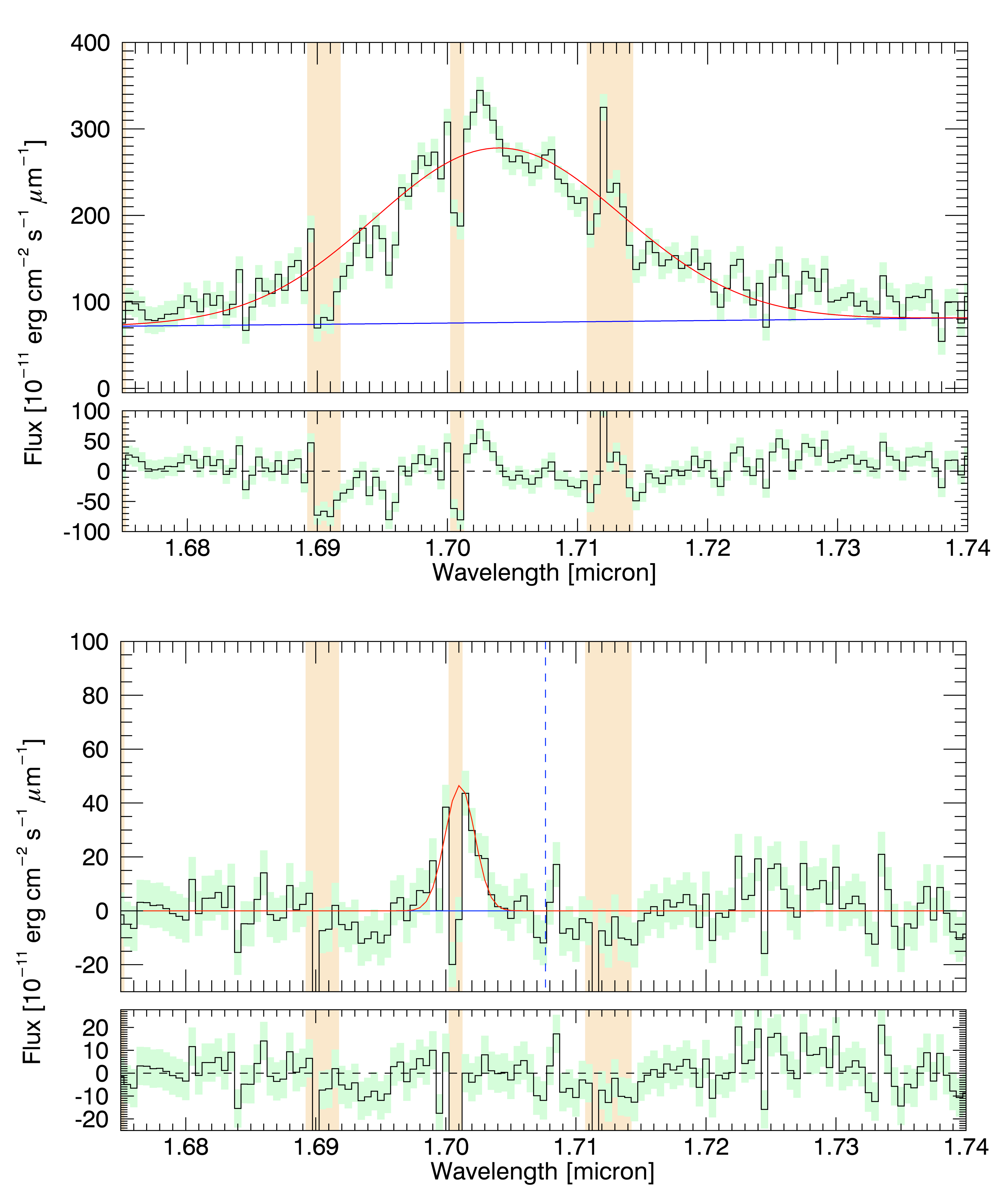}
	\end{center}
	\caption{\ha\ spectrum of XID2028. \textit{Upper panels:} the H band SINFONI spectrum of XID2028 in the spectral region of \ha, integrated in a region of $1"\times 1"$ around the QSO (i.e. the same region used to extract the spectrum shown in Fig.~\ref{Jband}). The observed spectrum is shown in black, while the single Gaussian fit is shown in red. The shaded regions mark the location of sky lines, that were excluded from the fit, while the residuals of the fit are shown below. A narrow \ha\ component is evident in the fit residuals, at the velocity of the broad component peak, corresponding to the systemic velocity of the system from optical spectroscopy. 
\textit{Lower panels:} \ha\ fitting residuals integrated over a region of $4\times4$ spaxels ($0.5"\times0.5"$) around the star forming region A shown in Fig.~\ref{Hintegrated}. The integrated spectrum is shown along with the fit to the narrow \ha\ line due to star formation in the host galaxy, shown in red. The line is detected at 9$\sigma$. The dashed blue line marks the expected position for [NII]$\lambda$6584: the line is non detected, providing an upper limit $log([NII]/H\alpha)<-1.1$, confirming that star formation is the excitation mechanism for the narrow \ha\ line. .
	}
	\label{Hband1D}
\end{figure}

The H+K SINFONI datacube is sampling the \ha\ line region, which in QSOs spectra is usually consisting of a broad component due to the AGN Broad Line Region (BLR, $FWHM\gtrsim3000\ km/s$), plus narrower components coming from the Narrow Line Region (NLR, $FWHM\gtrsim500\ km/s$) and  from the star formation in the host galaxy ($FWHM\lesssim300\ km/s$). The \ha\ spectrum of XID2028 integrated over the same $1"\times 1"$ area around the QSO nucleus used to extract the \oiiib\ spectrum is shown in Fig.~\ref{Hband1D}, upper panel. We fitted the integrated spectrum with a single broad Gaussian, finding a $FWHM=5300\ km/s$, consistent with the H$\beta$ Gaussian BLR component (see also Bongiorno et al. \citealp{bongiorno14}). 
Although a single Gaussian represents a good fit on the wings of the emission line, given the symmetric shape of the line (see Banerji et al \citealp{banerji12}), a weaker but significant narrow symmetric component is present in the fit residuals. The non detection of an additional asymmetric component, as observed for \hb, is probably due to the lower S/N in the data. In fact, a blue wing is clearly detected in deeper X-Shooter spectra of the same source (see Perna et al. \citealp{perna14}, Brusa et al. \citealp{brusa14}). The redshift $z=1.594$ of this narrow component, fitted with a single Gaussian profile ($FWHM=320\ km/s$), is consistent with the systemic redshift of the host galaxy, as measured by \oii\  in the optical Keck/DEIMOS spectrum (Brusa et al. \citealp{brusa10}). As for \hb, a velocity shift between the broad BLR component and the narrow component is present, as commonly observed in broad line AGNs (see e.g. Vanden Berk et al. \citealp{berk01}).

\begin{figure*}
	\begin{center}
		\includegraphics[width=1\textwidth]{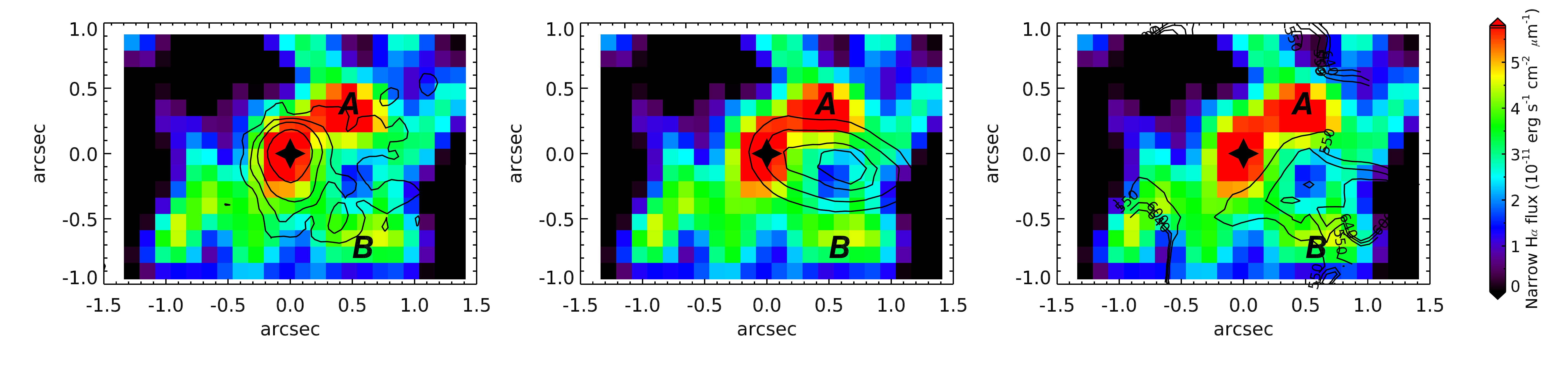}
	\end{center}
	\caption{Narrow H$\alpha$ map. The map is obtained integrating the single broad Gaussian \ha\ fit residuals on the spectral channels $1.7015<\lambda<1.7047\ \mu m$. In the \textit{left panel} the HST-ACS rest frame U band contours are superimposed in black (ACS level relative to the peak are 0.008, 0.015, 0.022, 0.05, 0.1, 0.5). The same pattern is obtained by these two independent tracers of star formation in the host galaxy, with two additional clumps of star formation (marked with A and B) elongated at the west of the QSO (marked with a star). In the \textit{central panel} the blue wing contours from Fig.~\ref{Jband}, tracing the outflow position, are plotted for comparison. A clear anti-correlation between the outflow location and the star formation tracers suggests that the outflowing material is sweeping the gas along the outflow core (`negative feedback'), while is compressing the gas at its edges inducing star formation at the locations marked as $A$ and $B$ on the map (`positive feedback'). The \textit{right panel} shows the $W_{40}$ lie width contours (i.e. the velocity width of the line that contains 80\% of the emission line flux such that $W_{40}=v_{50}-v_{10}$, where $v_{50}$ and $v_{10}$ are the velocities at the 50th and 10th percentiles, respectively; velocity levels 900, 1000, 1200 km/s) overplotted on the narrow \ha\ residuals. It can be seen how the shape of the \ha\ residuals, including the discontinuity between the central clump and the south west one, is anti correlated with regions of large line emission, $W_{40}>550\ km/s$, due to the outflowing gas.}
	\label{Hintegrated}
\end{figure*}

We map the spatial extent of the \ha\ narrow component by fitting in each spaxel of the datacube the single Gaussian derived from the integrated spectrum, keeping the width and centroid fixed and leaving the normalization free to vary. We show in Fig.~\ref{Hintegrated} the residual map obtained integrating the spectral channels corresponding to the narrow component ($1.7015<\lambda<1.7047\ \mu m$) in the datacube where the broad Gaussian fit has been subtracted. This narrow \ha\ map shows that 
the star formation in the host is not symmetrical distributed: most of the star formation activity is concentrated in the nucleus and in just two additional clumps, elongated to the west away from the QSO in two almost parallel branches around the location of the outflow (marked as A and B, although B is less significant in the \ha\ residual map).
The spectrum obtained integrating the residuals of the \ha\ fitting over a region of $4\times4$ spaxels ($0.5"\times0.5"$) around the star forming regions A is shown in Fig.~\ref{Hband1D}, lower panel, where the line is detected at $9\sigma$. The narrow H$\alpha$ line is also detected, although with a lower significance of 4$\sigma$, in the star forming region B.
We derive a limit of $log([NII]/H\alpha)<-1.1$ over the regions A and B. The narrow components of \hb\ and \oiiib\ are undetected across all the field of view, making impossible to place the two regions on classical AGN-star forming galaxies spectral diagnostic (BPT diagram, Baldwin et al. \citealp{baldwin81}). Nonetheless, the limit derived for [NII]/H$\alpha$ is already supporting a star forming origin of the line emission, as no known AGN shows such low values of this ratio (see Kauffmann et al. \citealp{kauffmann03} for the SDSS sample). 
Actually, high-z star forming galaxies do occupy the region of the BPT diagram with such low [NII]/Ha ratios (Steidel et al. \citealp{steidel14}), and often with [OIII]/Hb higher than the demarcation line between AGNs and star forming galaxies defined in the local Universe, due to a significant evolution of the ISM conditions with redshift (see e.g. Kewley et al. \citealp{kewley13a}, \citealp{kewley13b}).
For comparison, we measure $log([NII]/H\alpha)=-0.4$ for the narrow components in the nuclear region, fully consistent with AGN ionization. The different line ratios between the nucleus and the star forming regions indicate that scattered light from the QSO is not dominant at this wavelength. Therefore, assuming that all the narrow \ha\ is due to star formation, the total integrated narrow emission in the combined regions A and B corresponds to a total star formation rate $SFR\sim230\ M_{\odot}/yr$, using $E(B-V)=0.9$ as obtained by the SED decomposition fitting in Bongiorno et al. (\citealp{bongiorno12}). This is comparable with the value derived on the basis of the FIR emission (see Brusa et al. \citealp{brusa14}).

The same peculiar spatial pattern is also found in the HST-ACS F814W imaging, sampling the rest frame U band at the redshift of the source (Fig.~\ref{Hintegrated}, left panel). 
Although at least part of the U emission may be due to scattered light from the QSO (Zakamska et al. \citealp{zakamska06}), the U band is also sensitive to the light emitted by young, massive stars. Therefore, the similarity of the shape of the U-band image with that of the narrow \ha\ emission suggests that the UV continuum is also dominated by star formation that covers a horseshoe region around the outflow.
 
\begin{figure}
	\begin{center}
		\includegraphics[width=0.5\textwidth]{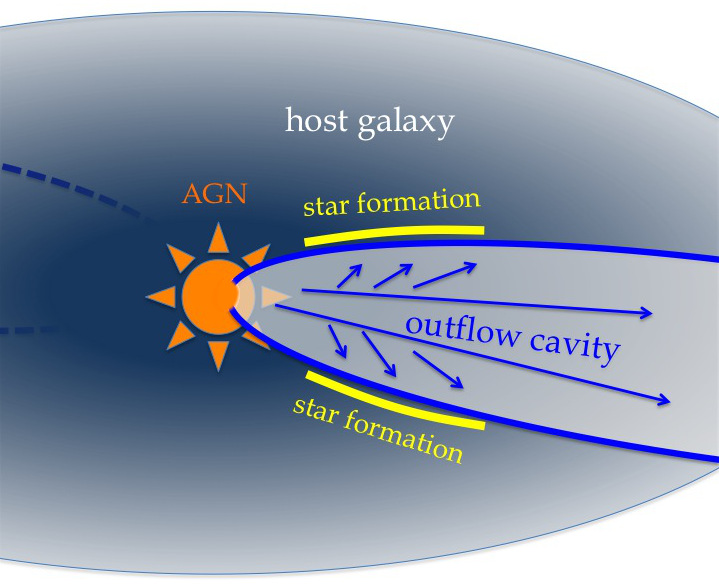}
	\end{center}
	\caption{Schematic view of the geometry of the system. The high velocity material is sweeping the gas suppressing star formation in a cavity along the core of the outflow (`negative feedback'), but also triggering star formation by outflow induced pressure at the edges (`positive feedback').}
	\label{model}
\end{figure}

 Interestingly, the outflow position is coincident with the cavity between these two star forming regions (Fig.~\ref{Hintegrated}, central and right panel), both in the narrow \ha\ residuals maps as well as in the ACS rest frame U image. The star formation activity on the host is therefore heavily suppressed in the core of the outflow, where the fast expanding gas is able to sweep away the gas needed to sustain the assemble of new stars. This represents a clear example of \textit{`negative feedback'} in action, showing the powerful outflow expelling most of the gas and quenching the star formation along its route in the host galaxy. 
On the other hand, we also find evidence that enhanced, triggered star forming activity is detected in the two off-center, elongated regions surrounding the bulk of the outflow, marked as $A$ and $B$ in Fig.~\ref{Hintegrated}. The causal connection between the outflows and the star forming regions is supported by the highly asymmetric shape of the star forming regions, that are extending out of the galaxy exactly along both edges of the outflow, while no other tail or star forming clump is detected in the rest of the host galaxy: the integrated flux of the narrow \ha\ line emission in the rest of the disk is less than half ($\sim40\%$) of the flux detected on regions A and B alone. A schematic view of the geometry of the system is shown in Fig.~\ref{model}.
If we instead assume that the outflow is just taking the path of least resistance through regions of lower gas density, we should also expect high velocity material in different directions. For example, region B appears to be too far ($1"\sim8$ kpc) to be able to collimate the outflow: in this alternative scenario, a prominent outflow developing towards the south in the gap between the nucleus and region B should also be present, contrary to the observations.
%


\section{Conclusions} \label{conclusions}

We have presented SINFONI observations of an obscured, radio-quiet X-ray selected QSO at $z=1.6$, XID2028, where indications of extended, fast outflows were available from previous X-Shooter data. Using an efficient criterion to select obscured AGNs at the maximum of their feedback activity based on the observed red colors (R-K$>4.5$) and high X-ray to optical flux ratio (X/O$>10$), the $z=1.6$ QSO was selected to be in the short lived blow-out phase. This galaxy is selected at the peak epoch of galaxy and black hole assembly, where we expect to have the maximum influence of feedback on the evolution of the host galaxies. Therefore, the results presented are particularly relevant in the broad picture of galaxy evolution, and an important complement to the more numerous observations of outflows in local galaxies.

Thanks to the spatial information of J band SINFONI near-IR integral field spectroscopy, we were able to map a remarkably extended (13 kpc) and fast ($v\geq1500\ km/s$) outflow in this source, fitting the \oiiib\ blueshifted line wing. Moreover, we studied the energetic of the outflow, finding a mass outflow rate $\dot M_{out}\sim300\ M_{\odot}/yr$ for the ionized component only, which would translate into a total outflow rate of $\dot M_{out}\gtrsim1000\ M_{\odot}/yr$ and a total kinetic power $P_{kin}\gtrsim 5.3\cdot 10^{44}\ erg\ s^{-1}$ once allowing for a contribution by the molecular and neutral phase. Although a more firm estimate of the outflowing mass will be obtained only with forthcoming mm observations (PdBI and ALMA), we stress that such an energetic outflow is hardly sustainable by star formation only, as discussed in Sect.~\ref{energetic}.

Finally, we were able to study the effects of such energetic outflow on the host galaxy, using both H+K band SINFONI observations sampling \ha\ and HST-ACS in the F814W filter. Both the narrow H$\alpha$ emission line map and the rest frame U band ACS imaging, tracing the star formation in the host galaxy, show that the outflow position lies in the center of a cavity in the star forming regions in the host galaxy. This is suggestive of a scenario in which the powerful outflow is removing the gas from the host galaxy (`negative feedback'), but also triggering star formation with the gas clouds compressed by the outflow-driven shock, which drives turbulent compression especially at the outflow edges (`positive feedback').  

Such `positive feedback' has been invoked in recent years to explain the correlation between AGN luminosities and nuclear star formation rates (Imanishi et al. \citealp{imanishi11}, Zinn et al. \citealp{zinn13}, Zubovas et al. \citealp{zubovas13}) as well as between black hole accretion rate and star formation rate in AGN (Silverman et al. \citealp{silverman09}, Mullaney et al. \citealp{mullaney12}). Moreover, one additional motivation for introducing AGN outflows (jets or winds) as a star formation trigger is that this mechanism introduces a time scale that is shorter than the gravitational time scale (Silk \& Norman \citealp{silk09}).  As AGN feedback is expected to be dominant at high redshift and rare at low redshift, models that are taking into account `positive' feedback as a second mode of star formation claim to naturally account for the observed evolution of sSFR (Silk \citealp{silk13}), elevated and more slowly varying at high redshift (see e.g. Stark et al. \citealp{stark13}), and the higher efficiency in star formation (Ishibashi \& Fabian \citealp{ishibashi12}, \citealp{ishibashi14}) observed in  sub-mm and ULIRG galaxies (e.g. Genzel et al. \citealp{genzel10}).  

Despite its possible importance in galaxy evolution, the few available observational evidences of such feedback in action were till now ascribed to an handful of extreme, powerful jets in radio-loud galaxies, in the local and high redshift universe. Moreover, the feedback-induced star formation has been usually found not in the AGN host galaxy, but in a companion satellite aligned along the radio axis. For example, Croft et al. \citep{croft06} and Elbaz et al. \citep{elbaz09} found jet induced star formation in companion galaxies of radio loud AGNs, while Kramer et al. \citep{klamer04} observations suggest that the radio jet from a $z=4.6$ QSO is triggering molecular cloud formation $\sim25\ kpc$ away from the host galaxy. On the other hand, Feain et al. \citep{feain07} found evidences of star formation aligned with the radio jet both in a QSO host galaxy at $z=0.3$ as well as in a companion, and Crockett et al. \citep{crockett12} revealed young stars near the filament of Centaurus A. Finally, Rodr{\'{\i}}guez-Zaur{\'{\i}}n et al. \citep{rz07} detected in a local, merging and radio loud ULIRG Super Star Clusters moving at high velocities (450 km/s) with respect to the local ambient gas, suggesting that they have been formed either in fast moving gas streams/tidal tails as part of the merger process or as a consequence of jet-induced star formation linked to the extended, diffuse radio emission in the halo of the galaxy. 

Therefore, XID2028 may represent the first direct detection of outflow induced star formation in a radio quiet AGN, as well as the first example of both types of feedback simultaneously at work in the same galaxy. The data presented demonstrate that both `positive' and `negative' AGN feedback are crucial ingredients to shape the evolution of galaxies, by regulating the star formation in the host and driving the BH-galaxy coevolution. Our results show that these mechanisms are in action not only in powerful radio galaxies with relativistic jets, but also in less extreme objects during an obscured QSO phase that is thought to be a common step in the evolutionary sequence of star forming galaxies. 


\acknowledgments

We are grateful to the ESO staff for their work and support. We also acknowledge the contribution of the entire COSMOS collaboration for making their data products publicly available. GC and AM acknowledge support from grant PRIN-INAF 2011 ``Black hole growth and AGN feedback through the cosmic time" and from grant PRIN-MIUR 2010-2011 ``The dark Universe and the cosmic evolution of baryons: from current surveys to Euclid". MB and MP acknowledge support FP7 Career Integration Grant "Supermassive Black Holes evolution through cosmic time: from current surveys to eROSITA-Euclid AGN synergies" (eEASy, CIG 321913). 



{\it Facilities:} \facility{VLT (SINFONI)}, \facility{HST (ACS)}




\end{document}